\documentclass[9pt]{article}
\usepackage{spconf,graphicx}
\usepackage{setspace}
\usepackage{enumitem}
\usepackage{amsmath}
\usepackage{xcolor}
\usepackage{hyperref}
\hypersetup{
    colorlinks=true,
    linkcolor=black,
    urlcolor=black,
    citecolor=black
}
\usepackage{booktabs}
\usepackage{multirow}
\usepackage{caption}
\usepackage{subcaption}
\usepackage{makecell}
\usepackage{textpos}
\usepackage[numbers, sort&compress]{natbib}
\usepackage{url}
\usepackage{amssymb}
\usepackage{pifont}
\title{Amphion: An Open-Source Audio, Music, and Speech Generation Toolkit}
\name{
\begin{tabular}{c}
  Xueyao Zhang$^{1,\star}$\thanks{$^{\star}$ Equal contribution.} \qquad  
  Liumeng Xue$^{1,\star}$ \qquad 
  Yicheng Gu$^{1,\star}$ \qquad 
  Yuancheng Wang$^{1,\star}$ \qquad
  Jiaqi Li$^{1,\star}$ \qquad \\
  Haorui He$^{3}$ \qquad 
  Chaoren Wang$^{1}$ \qquad 
  Songting Liu$^{3}$ \qquad 
  Xi Chen$^{1}$ \qquad 
  Junan Zhang$^{2}$ \qquad \\
  Zihao Fang$^{1}$ \qquad
  Haopeng Chen$^{1}$ \qquad 
  Tze Ying Tang$^{1}$ \qquad 
  Lexiao Zou$^{3}$ \qquad 
  Mingxuan Wang$^{1}$ \qquad \\
  Jun Han$^{1}$ \qquad 
  Kai Chen$^{2}$ \qquad 
  Haizhou Li$^{1}$ \qquad 
  Zhizheng Wu$^{1,2,3,\ddagger}$\thanks{$^{\ddagger}$ Correspondence to \textit{wuzhizheng@cuhk.edu.cn}.}  
\end{tabular}
}

\address{
\begin{tabular}{c}
  $^{1}$ The Chinese University of Hong Kong, Shenzhen, China\\
  $^{2}$ Shanghai AI Laboratory, Shanghai, China\\
  $^{3}$ Shenzhen Reseach Institute of Big Data, Shenzhen, China\\
\end{tabular}
}
\begin{document}
%
\maketitle
\begin{abstract}
Amphion is an open-source toolkit for \underline{A}udio, \underline{M}usic, and S\underline{p}eec\underline{h} Generat\underline{ion}, targeting to ease the way for junior researchers and engineers into these fields. It presents a unified framework that includes diverse generation tasks and models, with the added bonus of being easily extendable for new incorporation. The toolkit is designed with beginner-friendly workflows and pre-trained models, allowing both beginners and seasoned researchers to kick-start their projects with relative ease. 
The initial release of Amphion v0.1 supports a range of tasks including Text to Speech (TTS), Text to Audio (TTA), and Singing Voice Conversion (SVC), supplemented by essential components like data preprocessing, state-of-the-art vocoders, and evaluation metrics. This paper presents a high-level overview of Amphion. Amphion is open-sourced at \url{https://github.com/open-mmlab/Amphion}.
\end{abstract}


\section{Introduction}

The development of deep learning has greatly improved the performance of generative models. Leveraging these models has enabled researchers and practitioners to explore innovative possibilities, leading to notable breakthroughs across various fields, including computer vision and natural language processing. The potential in tasks related to audio, music, and speech generation has spurred the scientific community to actively publish new models and ideas \cite{fastspeech2, shen2023naturalspeech}. 

There is an increasing presence of both official and community-driven open-source repositories that replicate these models. \textbf{\textit{However, the quality of repositories varies}}, and \textbf{\textit{they are often scattered, focusing on specific papers}}. These scattered repositories introduce several obstacles to junior researchers or engineers who are new to the research area.
First, attempts to replicate an algorithm using different implementations or configurations can result in inconsistent model functionality or performance. 
Second, while many repositories focus on the model architectures, they often neglect crucial steps such as detailed data pre-processing, feature extraction, model training, and systematic evaluation. This lack of systematic guidance poses substantial challenges for beginners, who may have limited technical expertise and experience in training large-scale models. \textbf{\textit{In summary, the scattered nature of these repositories hampers efforts towards reproducible research and fair comparisons among models or algorithms.}} 

\begin{figure}[t]
    \centering
    \includegraphics[width=\columnwidth]{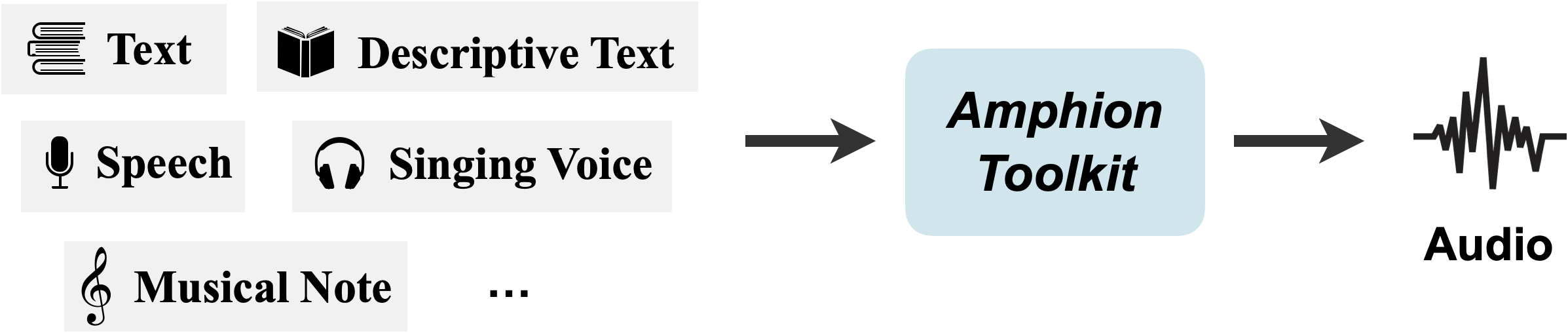}
    \caption{The north-star goal of Amphion toolkit: \textit{``Any to Audio"}.}\label{fig:northstar}
   \vspace{-10pt}
\end{figure}

Motivated by that, we introduce Amphion, an open-source platform dedicated to the north-star objective of ``Any to Audio'' (Figure~\ref{fig:northstar}). 
Amphion's features are summarized as:
\begin{itemize}[itemsep=0ex,leftmargin=3ex]
    \item \textbf{Unified Framework}: Amphion provides a unified framework for audio, music, and speech generation and evaluation. It is designed to be adaptable, flexible, and scalable, supporting the integration of new models.
    \item \textbf{Beginner-Friendly Workflow}: Amphion offers a beginner-friendly workflow with straightforward documentation and instructions. It establishes itself as an accessible one-stop research platform suitable for both novices and experienced researchers.
    \item \textbf{High-Quality Open Pre-trained Models}: To promote reproducible research, Amphion commits to releasing high-quality pre-trained models. In partner with industry, we aim to make large-scale pre-trained models widely available for various applications.
\end{itemize}
The Amphion v0.1 toolkit\footnote{\href{https://github.com/open-mmlab/Amphion}{https://github.com/open-mmlab/Amphion}}, now available under the MIT license, has supported a diverse array of generation tasks. This paper presents a high-level overview of the toolkit.

\section{The Amphion Framework}\label{sec:unified-framework}

The north-star goal of Amphion is to unify various audible waveform generation tasks. \textit{From the perspective of input, we formulate audio generation tasks into three categories},
\begin{enumerate}[itemsep=0ex,leftmargin=2ex]
    \item \textbf{Text to Waveform}: The input consists of discrete textual tokens, which strictly constrain the content of the output waveform. The representative tasks include Text to Speech (TTS) and Singing Voice Synthesis (SVS)\footnote{The inputs to SVS can also be musical tokens such as MIDI notes. They are also discrete and function like textual inputs.}.
    \item \textbf{Descriptive Text to Waveform}: The input consists of discrete textual tokens, which generally guide the content or style of the output waveform. The representative tasks include Text to Audio (TTA) and Text to Music (TTM).
    \item \textbf{Waveform to Waveform}: Both the input and output are continuous waveform signals. The representative tasks include Voice Conversion (VC), Singing Voic Conversion (SVC), Emotion Conversion (EC), Accent Conversion (AC), and Speech Translation (ST).
\end{enumerate}



\subsection{System Architecture Design}\label{sec:system-architecture-design}

\begin{figure}[t]
    \centering
    \includegraphics[width=0.95\columnwidth]{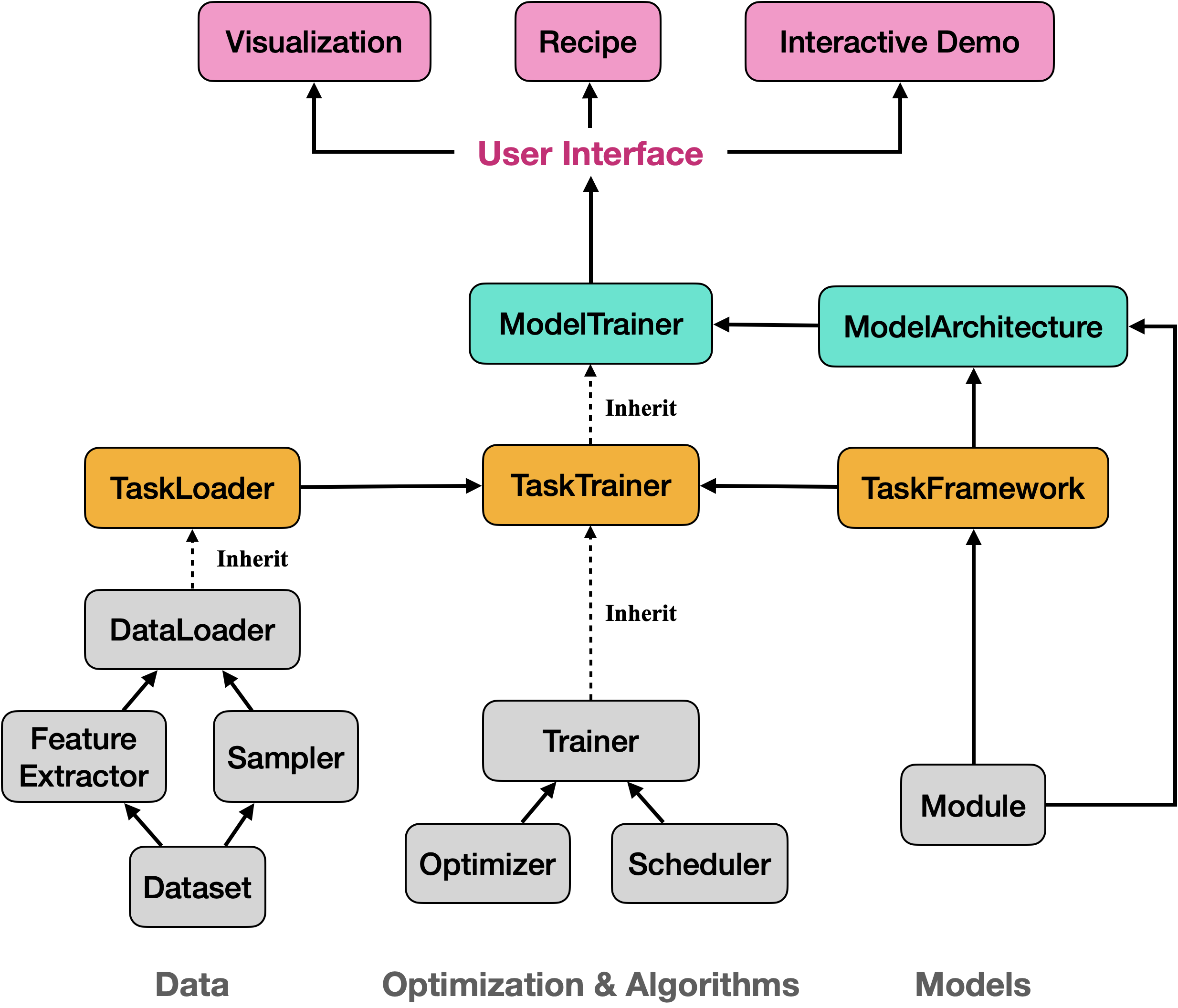}
    \caption{System architecture design of Amphion.
    }
    \label{fig:system-design}
     
\end{figure}

Amphion is designed to be a single framework supporting audio, music, and speech generation. 
Its system architecture design is presented in Figure~\ref{fig:system-design}. From the bottom up, 
\begin{enumerate}[itemsep=0ex,leftmargin=3ex]
    \item The data processing (\textit{Dataset}, \textit{Feature Extractor}, \textit{Sampler}, and \textit{DataLoader}), optimization algorithms (\textit{Optimizer}, \textit{Scheduler}, and \textit{Trainer}), and the common network modules (\textit{Module}) are shared building blocks for all the audio generation tasks.
    \item For each specific generation task, we unify its data/feature usage (\textit{TaskLoader}), task framework (\textit{TaskFramework}), and training pipeline (\textit{TaskTrainer}).
    \item Under each generation task, for every specific model, we specify its architecture (\textit{ModelArchitecture}) and training pipeline (\textit{ModelTrainer}).
    \item Finally, we provide a \textit{recipe} of each model for users. 
    On top of pre-trained models, we also offer interactive demos for users to explore. Amphion also features educative visualizations of machine learning models. Interested readers could refer to \cite{singvisio} for a detailed description.
\end{enumerate}

\subsection{Audio Generation Tasks Support}\label{sec:classic-generation-tasks}

Amphion v0.1 toolkit includes a representative from each of the three generation task categories (namely TTS, TTA, and SVC) for integration. This ensures that Amphion's framework can be conveniently adaptable to other audio generation tasks during future development. 
%


Specifically, the pipelines of different audio tasks are designed as follows:

\begin{itemize}[itemsep=0ex,leftmargin=2ex]
    \item \textbf{Text to Speech}: TTS aims to convert written text into spoken speech. Conventional \textit{multi-speaker TTS} are only trained with carefully-curated, few-speaker datasets and only produces speech from the speaker pool~\cite{fastspeech2,vits}. Recently,  \textit{zero-shot TTS} attracts more attentions from the research community. In addition to text, zero-shot TTS requires a reference audio as a prompt. By utilizing in-context learning techniques, it can imitate the timbre and speaking style of the reference audio~\cite{shen2023naturalspeech,valle}. 
    \item \textbf{Text to Audio}: TTA aims to generate sounds that are semantically in line with descriptions. It usually requires a pre-trained text encoder to capture the global information of the input descriptive text, and then utilizes an acoustic model, such as diffusion model~\cite{liu2023audioldm,wang2023audit}, to synthesize the acoustic features.
    \item \textbf{Singing Voice Conversion}: SVC aims to transform the voice of a singing signal into the voice of a target singer while preserving the lyrics and melody. To empower the reference speaker to sing the source audio, the main idea is to extract the speaker-specific representations from the reference, extract the speaker-agnostic representations (including semantic and prosody features) from the source, and then synthesize the converted features using acoustic models~\cite{multiple-contents-svc}.
\end{itemize}

\begin{table*}[t!]
    \centering
    \caption{Supported tasks, models, and metrics in Amphion v0.1.}
    \resizebox{0.9\textwidth}{!}{
        \small
        \begin{tabular}{ll||l}
        \toprule
        \multicolumn{2}{c}{\textbf{Tasks, Open Pre-trained Models, and Algorithms}} & \multicolumn{1}{c}{\textbf{Evaluation Metrics}} \\
        \midrule
        \begin{minipage}{0.32\textwidth}
          \fbox{\textbf{Text to Speech}}
          \begin{itemize}[itemsep=-1ex,leftmargin=2ex]
            \item \textbf{Transformer-based}: FastSpeech 2~\cite{fastspeech2}
            \item \textbf{Flow-based}: VITS~\cite{vits}
            \item \textbf{Diffusion-based}: NaturalSpeech 2~\cite{shen2023naturalspeech}
            \item \textbf{Autoregressive-based}: VALL-E~\cite{valle}
          \end{itemize}

          \fbox{\textbf{Singing Voice Conversion}}
          \begin{itemize}[itemsep=-1ex,leftmargin=2ex]
            \item \textbf{Transformer-based}: TransformerSVC
            \item \textbf{Flow-based}: \href{https://github.com/svc-develop-team/so-vits-svc}{VitsSVC}
            \item \textbf{Diffusion-based}: DiffWaveNetSVC~\cite{multiple-contents-svc}, DiffComoSVC~\cite{comosvc}
          \end{itemize}
        \fbox{\textbf{Text to Audio}}
          \begin{itemize}[itemsep=-1ex,leftmargin=2ex]
                \item AudioLDM~\cite{liu2023audioldm}, PicoAudio~\cite{picoaudio}
          \end{itemize}
        \end{minipage}
        & 
        \begin{minipage}{0.31\textwidth}
          \fbox{\textbf{Vocoder}}
          \begin{itemize}[itemsep=-1ex,leftmargin=2ex]
            \item \textbf{Autoregressive-based}: WaveNet~\cite{wavenet},  WaveRNN~\cite{wavernn}
            \item \textbf{Diffusion-based}: DiffWave~\cite{diffwave}
            \item \textbf{Flow-based}: WaveGlow~\cite{waveglow}
            \item \textbf{GAN-based}:
            \vspace{-2mm}
            \begin{itemize}[leftmargin=2ex,label={$\circ$}]
                \item \textbf{Generators}: MelGAN~\cite{melgan}, HiFi-GAN~\cite{hifigan}, NSF-HiFiGAN~\cite{sifigan}, BigVGAN~\cite{bigvgan}, APNet~\cite{apnet}
                \item \textbf{Discriminators}: MSD~\cite{melgan}, MPD~\cite{hifigan}, MRD~\cite{universalmelgan}, MS-STFTD~\cite{encodec}, MS-SB-CQTD~\cite{cqt, wavelet_cqt}
            \end{itemize}
          \end{itemize}
          \fbox{\textbf{Codec}}
          \begin{itemize}[itemsep=-1ex,leftmargin=2ex]
            \item FACodec~\cite{naturalspeech3}
          \end{itemize}
        \end{minipage}
        &
        \begin{minipage}{0.22\textwidth}
              \begin{itemize}[itemsep=0ex,leftmargin=2ex,rightmargin=\leftmargin]
                \item \textbf{F0 Modeling}: F0 Pearson Coefficients (FPC), Voiced/Unvoiced F1 Score (V/UV F1), etc.
                \item \textbf{Spectrogram Distortion}: PESQ, STOI, FAD, MCD, SI-SNR, SI-SDR
                \item \textbf{Intelligibility}: Word Error Rate (WER) and Character Error Rate (CER)
                \item \textbf{Speaker Similarity}: Cosine similarity, \href{https://github.com/resemble-ai/Resemblyzer}{Resemblyzer}, and WavLM.
              \end{itemize}
        \end{minipage}
        \\
        \bottomrule
        \end{tabular}    
    }
    \label{tab:supported-functions}
\end{table*}
 
Notably, most audio generation models usually adopt a two-stage generation process, where they generate some intermediate acoustic features (e.g. Mel Spectrogram) in the first stage, and then generate the final audible waveform using a vocoder or audio codec in the second stage. Motivated by that, Amphion v0.1 also integrates a variety of vocoder and audio codec models. 
A summary of Amphion's current supported tasks, models and algorithms is presented in Table~\ref{tab:supported-functions}.

\subsection{Open Pre-trained Models}\label{sec:open-pretrained-models}
Amphion has released a variety of models for TTS, TTA, SVC, and Vocoder. 

\begin{table}[h]
  \centering
    \caption{Supported pre-trained models in Amphion v0.1.}
  \resizebox{\columnwidth}{!}{
    \small
    \begin{tabular}{cccc}
    \toprule
    \textbf{Task} & \textbf{Model Architecture} & \textbf{\# Parameters} & \textbf{Training Set} \\
    \midrule
    \multirow{2}{*}{\makecell[c]{\\TTS}} 
    & \makecell[c]{VITS \cite{vits}} & 30M & HiFi-TTS~\cite{bakhturina2021hi} \\ \cmidrule(lr){2-4}
    & \makecell[c]{VALL-E \cite{valle}} & 250M & MLS~\cite{mls} \\ \cmidrule(lr){2-4}
    & \makecell[c]{NaturalSpeech2 \cite{shen2023naturalspeech}} & 201M & Libri-light~\cite{librilight} \\
    \midrule
    
    TTA & \makecell[c]{AudioLDM \cite{wang2023audit}} & 710M & AudioCaps~\cite{audiocaps} \\
    \midrule
    
    SVC & \makecell[c]{DiffWaveNetSVC \cite{multiple-contents-svc}} & 31M & Mixed (see Sec. 3.3) \\
    \midrule
    
    Codec & \makecell[c]{FACodec \cite{naturalspeech3}} & 140M & Libri-light~\cite{librilight} \\
    \midrule
    
    \multirow{2}{*}{\makecell[c]{\\Vocoder}} 
    & \makecell[c]{HiFi-GAN \cite{hifigan}} & 24M & LibriTTS~\cite{libritts} \\ \cmidrule(lr){2-4}
    & \makecell[c]{BigVGAN \cite{bigvgan}} & 112M & LibriTTS~\cite{libritts} \\
    \bottomrule  
  \end{tabular}  
  }
  \label{tab:pretrained_models}
\end{table}



\subsection{Comparison to Other Audio Toolkist}
We survey a list of representative open-source audio, music and speech generation toolkits in Table \ref{tab:opensource_toolkits}. Comparing these systems, we find that Amphion has a comprehensive audio generation task support, including general audio synthesis, music/sing synthesis, zero-shot and multi-speaker TTS. 
From informal comparisons, we also find Amphion to be more beginner-friendly with more online demos accessible than many other toolkits\footnote{By September 2024, Amphion has released 10 interactive demos on Hugging Face Spaces (\href{https://huggingface.co/amphion}{https://huggingface.co/amphion}) and OpenXLab (\href{https://openxlab.org.cn/usercenter/Amphion}{https://openxlab.org.cn/usercenter/Amphion})}.

\begin{table}[h!]
  \centering
    \caption{Representative open-source toolkits related to audio, music and speech generation (sorted alphabetically). Each repository name has a hyperlink to its web source.}
  \resizebox{\columnwidth}{!}{
    \begin{tabular}{c ccc ccc}
    \toprule
    \textbf{Toolkit}  & \textbf{Audio}  & \textbf{Music/Singing}  & \textbf{Zero-Shot TTS} & \textbf{Multi-Speaker TTS} \\ 
    \midrule
            \textbf{Amphion}  &  \ding{51} &  \ding{51} & \ding{51} & \ding{51} \\
            \midrule
            \href{https://github.com/facebookresearch/audiocraft/}{AudioCraft}                &  \ding{51}  &  \ding{51}  &    &   &  \\
            
            \href{https://github.com/suno-ai/bark}{Bark}            & \ding{51}  &  \ding{51}  &   & \ding{51}  &  \\

            \href{https://github.com/coqui-ai/TTS}{Coqui TTS}            &    &    &  &  \ding{51}  &  \\

            \href{https://github.com/netease-youdao/EmotiVoice}{EmotiVoice} & & & & \ding{51} \\
            \href{https://github.com/espnet/espnet}{ESPnet}                  &    & \ding{51}  &   & \ding{51} \\
            
            
            \href{https://github.com/CSTR-Edinburgh/merlin}{Merlin} &    &    &  &  \ding{51}  &  \\

            \href{https://github.com/babysor/MockingBird}{Mocking Bird}             &    &    & \ding{51}  &   &  \\
            
            \href{https://github.com/SJTMusicTeam/Muskits}{Muskits}                 &    & \ding{51}   &    &   &  \\
            \href{https://github.com/microsoft/muzic}{Muzic}                   &    & \ding{51}  &    &   \\

            \href{https://github.com/myshell-ai/OpenVoice}{OpenVoice} & & & \ding{51} \\
           \href{https://github.com/PaddlePaddle/PaddleSpeech}{PaddleSpeech}            &    &  \ding{51}  & \ding{51}   & \ding{51}  & \\
           
            \href{https://github.com/svc-develop-team/so-vits-svc}{SoftVC VITS}             &    & \ding{51}  &    &   & \\
            \href{https://github.com/speechbrain/speechbrain}{SpeechBrain}             &    &    &  & \ding{51}   &  \\

            \href{https://github.com/neonbjb/tortoise-tts/}{TorToiSe}    &    &    &  &  \ding{51}  & \\

            \href{https://github.com/wenet-e2e/wetts}{WeTTS}                   &    &    &   &  \ding{51}  & \\
            
            \bottomrule 
  \end{tabular}  
  }
    \vspace{-5mm}
  \label{tab:opensource_toolkits}
\end{table}

\section{Experiments}\label{sec:experiments}
In this section, we compare the performance of models trained with Amphion v0.1 framework with public open repositories or results from original academic papers.
We also briefly mention the training configurations of the pretrained models in Amphion v0.1. We recommend interested readers to find more information in our repository.

We use both objective and subjective evaluations to evaluate. The objective evaluation metrics will be specified in each task. 
For subjective scores, including the naturalness Mean Opinion Score (MOS) and the Similarity Mean Opinion Score (SMOS), 10 listeners experienced in this field are requested to grade  from 1 (``Bad'') to 5 (``Excellent'') on 10 randomly selected sample audios on each condition, to assess each audio's overall quality and similarity to the reference speaker.


\subsection{Text to Speech}
\vspace{-3mm}
\begin{table}[!ht]
  \centering
    \caption{Evaluation results of multi-speaker TTS in Amphion v0.1.}
  \footnotesize
 \resizebox{\columnwidth}{!}{
    \begin{tabular}{c ccc ccc}
    \toprule
     \textbf{Systems}    & \textbf{CER $\downarrow$}  & \textbf{WER $\downarrow$}  & \textbf{FAD $\downarrow$}  & \textbf{MOS $\uparrow$ }\\                                   
    \midrule
    \makecell[c]{{Coqui TTS} (VITS)}      &    0.06 &  0.12   &   0.54 & 3.69  \\
   \makecell[c]{ {SpeechBrain} (FastSpeech 2) }   &    0.06 &  0.11   &  1.71 & 3.54  \\
    \makecell[c]{{TorToiSe TTS}}   &    0.05 &  0.09   &  1.90 & 3.61   \\
   \makecell[c]{ {ESPnet} (VITS)}   &    0.07 &  0.11   &  1.28 & 3.57  \\ \midrule
   \makecell[c]{ Amphion v0.1 (VITS)}         &    0.06 &  0.10   &   0.84 & 3.61 \\
    \bottomrule
  \end{tabular} 
 } 
  \label{tab:amphion_tts_res}
\end{table}


\subsubsection{Results of Multi-Speaker TTS}

For multi-speaker TTS, we compare Amphion v0.1 with other four popular open-source speech synthesis toolkits, including Coqui TTS\footnote{\href{https://github.com/coqui-ai/TTS}{https://github.com/coqui-ai/TTS} }, SpeechBrain\footnote{\href{https://github.com/speechbrain/speechbrain}{https://github.com/speechbrain/speechbrain} } , TorToiSe\footnote{\href{https://github.com/neonbjb/tortoise-tts}{https://github.com/neonbjb/tortoise-tts} } , and ESPnet\footnote{\href{https://github.com/espnet/espnet}{https://github.com/espnet/espnet} } .
For each open-source system, we select the best-performing multi-speaker model for the comparison. VITS~\cite{vits} is selected for Coqui TTS, ESPnet and Amphion, and FastSpeech 2~\cite{fastspeech2} is selected in SpeechBrain, and the TorToiSe TTS model from its repository. 
We evaluate on 100 text transcriptions and then generate the corresponding speech using each system.
The results are shown in Table~\ref{tab:amphion_tts_res}, which shows that the VITS multi-speaker TTS model released in Amphion v0.1 is comparable to existing open-source systems.

\subsubsection{Results of Zero-Shot TTS}
\vspace{-3mm}
\begin{table}[!ht]
    \centering
    \caption{Continuation evaluation results of VALL-E zero-shot TTS system in Amphion v0.1.}
    \resizebox{\columnwidth}{!}{
    \begin{tabular}{ccccc}
    \toprule
        \textbf{Systems} & \textbf{Training Dataset} & \textbf{Test Dataset} & \textbf{SIM-O $\uparrow$} & \textbf{WER $\downarrow$} \\ \midrule
        \multirow{2}{*}{Proprietary (VALL-E)}  & \multirow{2}{*}{Librilight} & Librispeech & \multirow{2}{*}{0.51} & \multirow{2}{*}{0.038} \\
        &&test-clean (4-10s) && \\ \midrule
        \multirow{2}{*}{Amphion v0.1(VALL-E)} & \multirow{2}{*}{MLS (10-20s)} & Librispeech & \multirow{2}{*}{0.51} & \multirow{2}{*}{0.034} \\
        &&test-clean (10-20s)&& \\ \bottomrule
    \end{tabular}
    }
\end{table}
For zero-shot TTS, we compare the VALL-E~\cite{valle} model released in Amphion v0.1 with the proprietary model results from the official paper. 
We test the objective speaker similarity score SIM-O, and WER (Word Error Rate), using the same evaluation tools as the official paper~\cite{valle}.
For SIM-O, we use the WavLM-TDNN \footnote{\href{https://github.com/microsoft/UniSpeech/tree/main/downstreams/
speaker verification}{https://github.com/microsoft/UniSpeech/tree/main/downstreams/
speaker verification}} model to extract speaker verification features, and use the Hubert-large~\cite{hubert} ASR system to trascribe the speech. 
Since our training set only contains 10-20s speech, we test results on a matched duration of LibriSpeech test-clean (10-20s). 
We test in a continuation setting following the VALL-E paper~\cite{valle}, where the model is given a 3-second prefix from the ground-truth utterance and asked to continue the speech.
The results show that in a matched train-test duration scenario, our model achieves a SIM-O and WER result comparable to the official paper.

For model training, we use the MLS~\cite{mls} dataset containing 45k hours of speech, which is close to the 60k hours of Libri-Light data for the official model. 
Notably, our released VALL-E model has utilized more training data than existing open-source models\footnote{\href{https://github.com/Plachtaa/VALL-E-X}{https://github.com/Plachtaa/VALL-E-X} }, which are typically trained on hundreds of hours of data.

\subsection{Text to Audio}
\vspace{-3mm}
\begin{table}[h]
  \centering
\caption{Evaluation results of Text to Audio in Amphion v0.1.}
  \footnotesize
  \scalebox{0.9}{
  \begin{tabular}{c cc ccc}
    \toprule
     \textbf{Systems}   & \textbf{FD $\downarrow$}  & \textbf{IS $\uparrow$} & \textbf{KL $\downarrow$}\\

    \midrule
    
     Text-to-sound-synthesis (Diffsound) & 47.68 & 4.01 & 2.52 \\
    AudioLDM (AudioLDM) & 27.12& 7.51 &1.86 \\
    \midrule
    \makecell[c]{Amphion v0.1 (AudioLDM)} &20.47 &8.78 &1.44 \\
    \bottomrule
  \end{tabular}  
  }
  \label{tab:amphion_tta_res}   
  \vspace{-3mm}
\end{table}

    

We compare the TTA models in different repositories: The Text-to-sound-synthesis\footnote{\href{https://github.com/yangdongchao/Text-to-sound-Synthesis}{https://github.com/yangdongchao/Text-to-sound-Synthesis}} repository with the DiffSound~\cite{yang2023diffsound} model, the official AudioLDM~\cite{liu2023audioldm} repository\footnote{\href{https://github.com/haoheliu/AudioLDM}{https://github.com/haoheliu/AudioLDM}}, and the reproduced AudioLDM model using Amphion's infrastructure.

To evaluate our text-to-audio model, we use inception score (IS), Fréchet Distance (FD), and Kullback–Leibler Divergence (KL). 
FD, IS, and KL are based on the state-of-the-art audio classification model PANNs~\cite{kong2020panns}. 
We use the test set of AudioCaps as our test set. The evaluation results of Amphion v0.1 TTA are shown in Table~\ref{tab:amphion_tta_res}. 
The results demonstrate that the Amphion v0.1 TTA system achieves superior results than existing open-source models.

\subsection{Singing Voice Conversion}

We compare the SVC system in 
Amphion v0.1 (\cite{multiple-contents-svc}) with 
the SoftVC~\footnote{\href{https://github.com/bshall/soft-vc}{https://github.com/bshall/soft-vc }} toolkit. 
To train our SVC model, we utilize a wide range of datasets: Opencpop~\cite{opencpop}, SVCC\footnote{\label{foot:example}\href{http://vc-challenge.org/}{http://vc-challenge.org/}} training data, VCTK\footnote{\href{https://huggingface.co/datasets/CSTR-Edinburgh/vctk}{https://huggingface.co/datasets/CSTR-Edinburgh/vctk}}, OpenSinger~\cite{multisinger}, and M4Singer~\cite{m4singer}. There are 83.1 hours of speech and 87.2 hours of singing data in total.

To evaluate the models, we adopt the in-domain evaluation task of the Singing Voice Conversion Challenge (SVCC) 2023\textsuperscript{13} with 48 singing utterances under test.
The task is to convert each singing utterance into two target singers (one male and one female). 
Results show that Amphion v0.1 SVC model owns better performance in both naturalness and speaker similarity than SoftVC, and narrowing the gap to ground truth utterances.

\begin{table}[h]
\centering
\small
  \centering
    \caption{Evaluation results of Singing Voice Conversion in Amphion v0.1.}
  \resizebox{0.8\columnwidth}{!}{
    \small
    \begin{tabular}{c ccc ccc}
    \toprule
     \textbf{Systems}  & \textbf{MOS $\uparrow$}  & \textbf{SMOS $\uparrow$ }\\                              
    \midrule
    \makecell[c]{Ground truth}    & 4.67   & 3.96 \\
    \makecell[c]{SoftVC (VITS)}     & 2.98   & 3.43 \\
    \midrule
    \makecell[c]{Amphion v0.1 (DiffWaveNetSVC)}         & 3.52    & 3.69 \\
    \bottomrule
  \end{tabular}  
  }

  \label{tab:amphion_svc_res}  
  \vspace{-10pt}
\end{table}

\begin{table}[h]
\footnotesize
  \centering\vspace{-2mm}
    \caption{Evaluation results of Vocoder in Amphion v0.1.}
  \resizebox{\columnwidth}{!}{
  \begin{tabular}{c ccc cc}
    \toprule
     \textbf{Systems}    & \textbf{PESQ $\uparrow$}  & \textbf{M-STFT $\downarrow$}  & \textbf{F0RMSE $\downarrow$} & \textbf{FPC $\uparrow$}  \\                                          
    \midrule
   \makecell[c]{  Official (HiFi-GAN)}    &    3.43 &  1.98   &  177 & 0.88  \\
   \makecell[c]{  ESPnet (HiFi-GAN) }  &    3.55 &  1.12   &  188 & 0.86  \\ \midrule
   \makecell[c]{  Amphion v0.1 (HiFi-GAN) }        &    3.55 &  1.09   &  188 & 0.88 \\
    \bottomrule
  \end{tabular}  
  }
  \label{tab:amphion_vocoder_res}
  \vspace{-5mm}

\end{table}

\subsection{Vocoder}

We compare the Amphion v0.1 Vocoder with the two widely used open-source HiFi-GAN checkpoints. One is the official HiFi-GAN repository\footnote{\href{https://github.com/jik876/hifi-gan}{https://github.com/jik876/hifi-gan} }; the other is from ESPnet\footnote{\href{https://github.com/kan-bayashi/ParallelWaveGAN}{https://github.com/kan-bayashi/ParallelWaveGAN} }. All of the checkpoints are trained on around 600 hours of speech data. The whole evaluation set and the test set of LibriTTS are used for evaluation, with a total of 20306 utterances. Objective evaluations are conducted with M-STFT, PESQ, F0RMSE, and FPC metrics. The results are illustrated in table~\ref{tab:amphion_vocoder_res}. With the assistance of additional guidance from Time-Frequency Representation-based Discriminators~\cite{cqt,wavelet_cqt}, the Amphion v0.1 HiFi-GAN achieves superior performance in spectrogram reconstruction and F0 modeling.

\section{Conclusion}
This paper presented Amphion, an open-source toolkit dedicated to audio, music, and speech generation. 
Amphion's primary objective is to facilitate reproducible research and serve as a stepping stone for junior researchers and engineers entering the field of audio, music, and speech generation. 
Since the release of Amphion in November 2023, Amphion has received more than 4,300 stars on GitHub and received a significant number of pull requests and feedback.
For future plans, Amphion is releasing a few large-scale datasets~\cite{emilia} in the area of audio, music and speech generation.
Also, we plan to partner with industry for releasing large-scale and production-oriented pre-trained models. 





\bibliographystyle{IEEEbib}
\bibliography{reference}

\end{document}